\documentclass[showpacs,aps,physrev,superscriptaddress, twocolumn, nofootinbib]{revtex4-2}
\usepackage{graphicx}
\usepackage{amsthm,amssymb,amsmath,braket,mathdots}
\usepackage[dvipsnames]{xcolor} 
\usepackage{float}
\definecolor{lightgreen}{RGB}{0, 200, 0}
\usepackage[colorlinks=true, urlcolor=Blue, linkcolor=cyan, citecolor=lightgreen]{hyperref} %%make hyperlinks in colors %%default hyperref sets: links (like table of contents) → red, citations → green, URLs → magenta
\usepackage{soul} %%for strikethrough \st{}
\usepackage{enumitem} %%for numerating items

\usepackage{mathtools}

\DeclarePairedDelimiter\floor{\lfloor}{\rfloor}

\newcommand{\affA}{Weierstrass Institute for Applied Analysis and Stochastics, Anton-Wilhelm-Amo-Stra{\ss}e 39, Berlin 10117, Germany}
\newcommand{\affB}{Japan Agency for Marine-Earth Science and Technology, Yokohama 236-0001, Japan}
\newcommand{\affC}{RIES/Department of Mathematics, Hokkaido University, Hokkaido 060-0812, Japan}
\newcommand{\affD}{Department of Complexity Science and Engineerings, Graduate School of Frontier Sciences, The University of Tokyo, Chiba 277-8561, Japan}

\begin{document}

\title{Basin Metamorphosis in Coupled Phase Oscillators 
} 

\author{Jin Yan}
\email{jin.yan@wias-berlin.de}
\affiliation{\affA}

\author{Ayumi Ozawa}
\email{ozawaa@jamstec.go.jp}
\affiliation{\affB}

\author{Hiroshi Kori}
\email{kori@k.u-tokyo.ac.jp}
\affiliation{\affD}
 
\author{Yuzuru Sato}
\email{ysato@math.sci.hokudai.ac.jp}
\affiliation{\affC}

\begin{abstract}
We investigate the global basin structure of twisted states in nearest-neighbor coupled phase oscillators with a common phase shift $\alpha$. 
As $\alpha$ increases, basin boundaries become progressively more complex, with their fractal dimension growing toward that of the full ambient phase space. We conjecture that the basins eventually become riddled-like as the system approaches the limit $\alpha\to \frac{\pi}{2}$, where the dynamics becomes  volume-preserving.  
We characterize the transient dynamics via the stabilization time of the winding number and demonstrate that it grows with system size. The scaling accelerates at larger phase shifts, transitioning from logarithmic to power-law behavior. We further analyze the dynamical origin of these long transients. 
Our results demonstrate how a single phase-shift governs fractal basin complexity and provide new insights into the global geometry and transient dynamics of multistable, yet non-chaotic, coupled phase oscillators.
\end{abstract}

\maketitle

\textit{1.~Introduction.---}
Basins of attraction provide a global view of the behavior of a dynamical system and play an important role in science and technology. They determine which long-term state a system will reach from a given set of initial conditions and thus reflect robustness, reliability and functionality. 
In engineered systems \cite{strogatz2024nonlinear}, brains and neural circuits \cite{bollt2023fractal, le2025asymmetric}, climate dynamics \cite{wunderling2020basin, dijkstra2013nonlinear}, power grids \cite{pikovsky1985universal, eaves2025nonlinear} and ecology \cite{feudel2023rate, may2001stability}, a desired state is relevant only when its basin of attraction is sufficiently large and smooth; otherwise, small perturbations can drive the system to an unintended state. 
In many nonlinear systems, however, basins are highly intricate, exhibiting fractal or even riddled structures \cite{alexander1992riddled, ashwin2000riddling, aguirre2009fractal, medeiros2018boundaries, dos2020basin, czajkowski2024riddled}. Such sensitivity fundamentally limits the predictability, reproducibility, and controllability of the resulting phenomena. 
In low-dimensional dynamical systems, fractal basin structures have been extensively studied \cite{alexander1992riddled,ott1993scaling}, particularly in relation to bifurcations and chaotic saddles. 

In coupled oscillator systems, basin geometry remains largely unexplored \cite{zhang2021basins}. Most previous works has adopted a statistical-physics viewpoint, focusing on collective behavior, ensemble averages and asymptotic states in the thermodynamic limit. This framework has been remarkably successful in describing synchronization transitions and macroscopic phases. However, it naturally obscures fine geometric structures in phase space. 
As a result, the fractal structure of basins of attraction has rarely been examined in the context of coupled oscillators, even though the transition of basins from smooth to fractal and eventually to riddled as a system parameter varies is a well-known phenomenon in nonlinear dynamics. 
What is particularly surprising is that fractal basin boundaries emerge in a paradigmatic system of coupled phase oscillators in the absence of chaos. 
In this work, we uncover the fine structure of basin boundaries in a minimal model of coupled phase oscillators. By examining low-dimensional slices of the high-dimensional phase space, we explicitly visualize fractal basin boundaries associated with different periodic attractors. The observed fractal nature persists for generic local slices. 
In addition, we connect this geometric complexity to the system's transient dynamics, and show numerically that the transient-time scaling with the system size is strongly influenced by fractal nature of basins, revealing features that remain invisible in asymptotic or averaged descriptions. 

We consider a ring of $N$ nearest-neighbor coupled phase oscillators \cite{sakaguchi1986soluble, kuramoto2002coexistence, wiley2006size, zhang2021basins} with a common phase shift $\alpha \in \left[ 0, \frac{\pi}{2}\right)$ 
\begin{equation}
\dot{\theta}_j = \sin (\theta_{j-1}-\theta_j+\alpha) + \sin (\theta_{j+1}-\theta_j+\alpha), 
\label{eq:sys-theta}
\end{equation}
where each phase $\theta_j \in [0, 2\pi)$, $j=1, 2, ..., N$, and with periodic boundary conditions. 
In this phase-reduced description of coupled limit-cycle oscillators \cite{sakaguchi1986soluble, kuramoto2002coexistence}, the phase-shift parameter $\alpha$ appears naturally through the reduction, breaking the symmetry of the coupling function while connecting the gradient system ($\alpha=0$) to the Hamiltonian system ($\alpha=\frac{\pi}{2}$)
\cite{rosenau2005phase, ahnert2008traveling, smirnov2018solitary}. Denoting the neighboring phase difference as $\psi_j(t) := \theta_{j+1}(t) - \theta_j(t) \in [-\pi, \pi)$, the system \eqref{eq:sys-theta} can be written as 
\begin{equation}
\begin{split}
\dot{\psi}_j =& \sin (\psi_{j+1}+\alpha) - \sin (\psi_j + \alpha)\\
&+  \sin (\psi_{j-1} - \alpha) - \sin (\psi_j - \alpha). 
\label{eq:sys-psi}
\end{split}
\end{equation}
It is known that at $\alpha=0$, the only attractors of the system are the so-called $q$-twisted states \cite{wiley2006size, zhang2021basins}, characterized by 
\begin{equation}
\psi_j = \Delta_q := \frac{2\pi q}{N}, \quad q = 0, \pm 1, \pm 2, ..., \pm \floor*{\frac{N}{4}}.  
\label{eq:qstate}
\end{equation}
In fact, they are stable also with  $\alpha>0$: the Jacobian spectrum of the system \eqref{eq:sys-theta} at a $q$-twisted state is given by 
$\lambda_k = 2\cos \alpha \cos \Delta_q (\cos k - 1) - i(2\sin\alpha \sin \Delta_q \sin k)$, 
with $k = \frac{2\pi m}{N}$, $m = 0, 1, ..., N-1$ \cite{kim2026multistability}. The Jacobian spectrum characterizes the local convergence of trajectories toward an attractor. 
The neutral mode $\lambda_0 = 0$ corresponds to a global translational symmetry $\boldsymbol{\theta} \to \boldsymbol{\theta} + \boldsymbol{c}$ of the solution, where $\boldsymbol{\theta} := (\theta_1, ..., \theta_N)$ and $\boldsymbol{c} \in \mathbb{R}^N$ is a constant vector. For all other modes $k>0$, Re$(\lambda_k) < 0$ for any $\alpha \in \left[ 0, \frac{\pi}{2}\right)$ and $|\Delta_q| < \frac{\pi}{2}$ (or equivalently $|q| < \frac{N}{4}$),  
the $q$-twisted states defined in Eq.~\eqref{eq:qstate} are stable. 
At $\alpha=0$, all the imaginary parts vanish; at $\alpha = \frac{\pi}{2}$, all the real parts vanish and the $q$-twisted states become marginally stable. 
Note that at $\alpha=0$ the $q$-twisted states are stationary ($\dot{\theta}_j = 0$), while for $\alpha>0$ they are rotating with a constant frequency $\dot{\theta}_j = \Omega_{\alpha, q} := 2\sin \alpha \cos \Delta_q$. 
When $\alpha=\frac{\pi}{2}$, 
Eq.~(2) yields a Hamiltonian system with Hamiltonian $H=\sum_{j=1}^N \sin(\psi_j)$ showing solitons and kinks in the continuum limit \cite{rosenau2005phase, ahnert2008traveling}.

\textit{2.~Fractal basin boundaries.---} 
At $\alpha=0$, as described in \cite{zhang2021basins}, each basin of the $q$-twisted states resembles an {\it octopus}, with a central {\it head} and extended {\it tentacles}, yet most part of the basin volume is contained in the tentacles. 
Being a gradient system, the dynamics is fully determined by its energy function and convergence to an attractor is rather fast and straightforward \cite{groisman2025size}. 
When a common phase shift $\alpha>0$ is introduced, however, the convergence slows down and in the limit $\alpha\to \frac{\pi}{2}$, the dynamics is volume-preserving. 
These observations raise two interrelated questions: how does the basin structure evolve as $\alpha$ varies, and how is this evolution reflected in the transient dynamics?

\begin{figure}[t]
\centering 
\includegraphics[width=\linewidth]{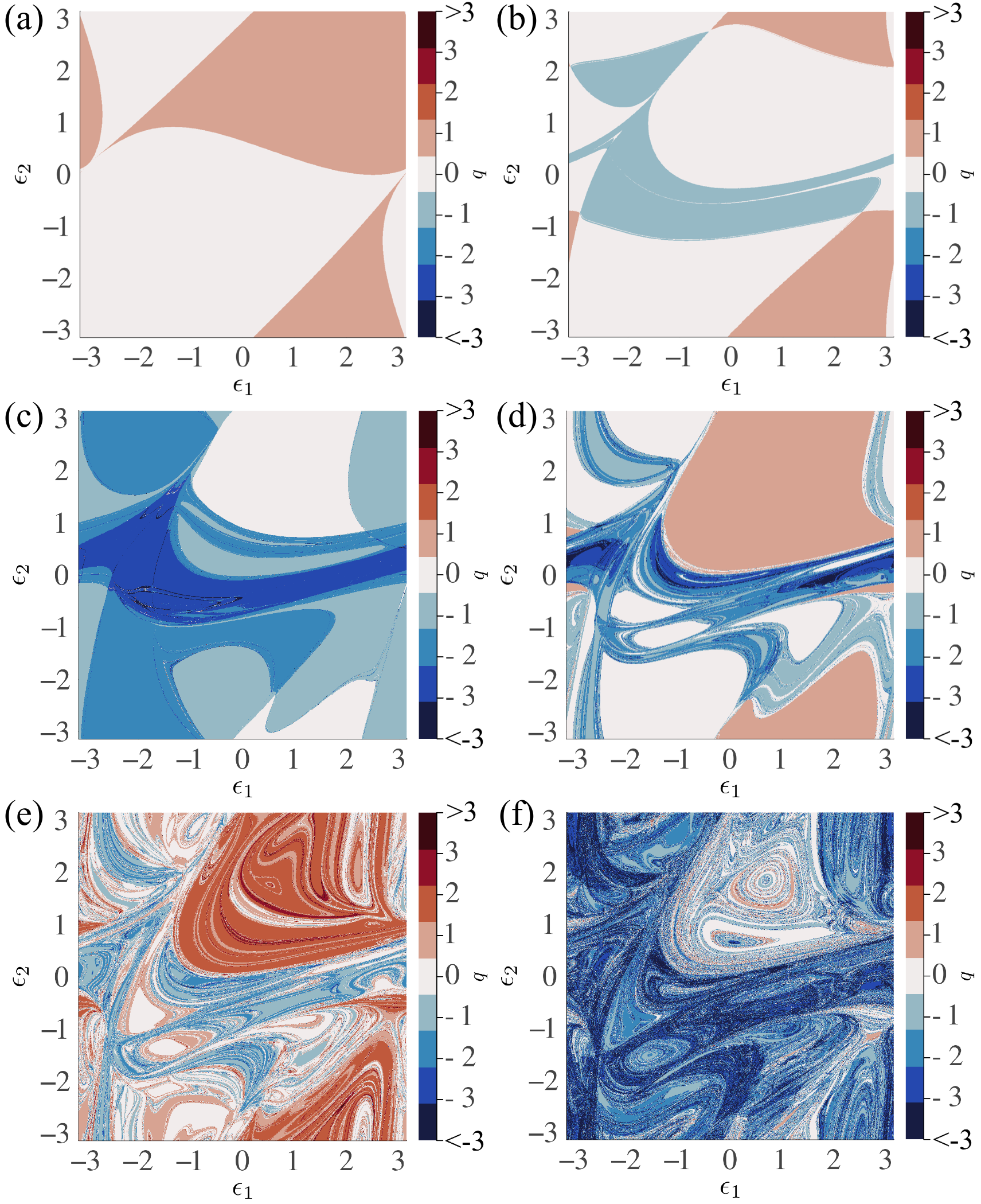} 
\caption{Basin metamorphosis of the $q$-twisted states on an $(\epsilon_1, \epsilon_2)$-slice centered at a random point $\boldsymbol{\theta}_b\in [-\pi, \pi)^N$: (a) $\alpha = 0$, (b) $\alpha=0.3\pi$, (c) $\alpha=0.36\pi$, (d) $\alpha=0.4\pi$, (e) $\alpha=0.42\pi$, and (f) $\alpha=0.44\pi$. Each point represents a phase-space state with $\boldsymbol{\theta}_b + (\epsilon_1, \epsilon_2, 0, ..., 0)$ and $N=80$. Resolution of each plot is $1000\times 1000$. The numerical scheme is specified in \cite{note1}.}
\label{fig:basins-delpi}
\end{figure}

For simplification, we consider a two-dimensional subspace as follows: we choose a base point in the phase space $\boldsymbol{\theta}_b \in [0, 2\pi)^N$, and consider a perturbation $\boldsymbol{\epsilon} = (\epsilon_1, \epsilon_2, 0, ..., 0)\in [-\pi, \pi)^N$, where $(\epsilon_1, \epsilon_2) \in [-\pi, \pi)^2$. 
This particular subspace with a fixed base point is chosen because, due to nearest-neighbor coupling, perturbations along these two coordinates give the strongest dynamical effects among all possible oscillator pairs. 
The basin structure thus can be projected onto the $(\epsilon_1, \epsilon_2)$-slice. 
This perturbation should be considered small for large $N$. 
However, even in this two-dimensional subspace, the basin structure becomes remarkably complex as $\alpha$ increases. 
Figure \ref{fig:basins-delpi} shows the basins of the $q$-twisted states on the same slice centered at a random phase-space point: at $\alpha=0$, the basins appear fragmented \cite{zhang2021basins}, and in this particular slice, only the basins of the $q=0$ and $q=1$ twisted states are observable. As $\alpha$ increases, additional basins emerge; for $\alpha > 0.4\pi$, all twisted states with $|q|\leq 3$ are observable, at the same time basin boundaries become increasingly fractal. 
We note that the system becomes numerically delicate as the basins develop fractality, and resolving finer details requires higher numerical accuracy in the integration scheme \cite{note1}. 

\begin{figure}[H]
\centering 
\includegraphics[width=0.6\linewidth]{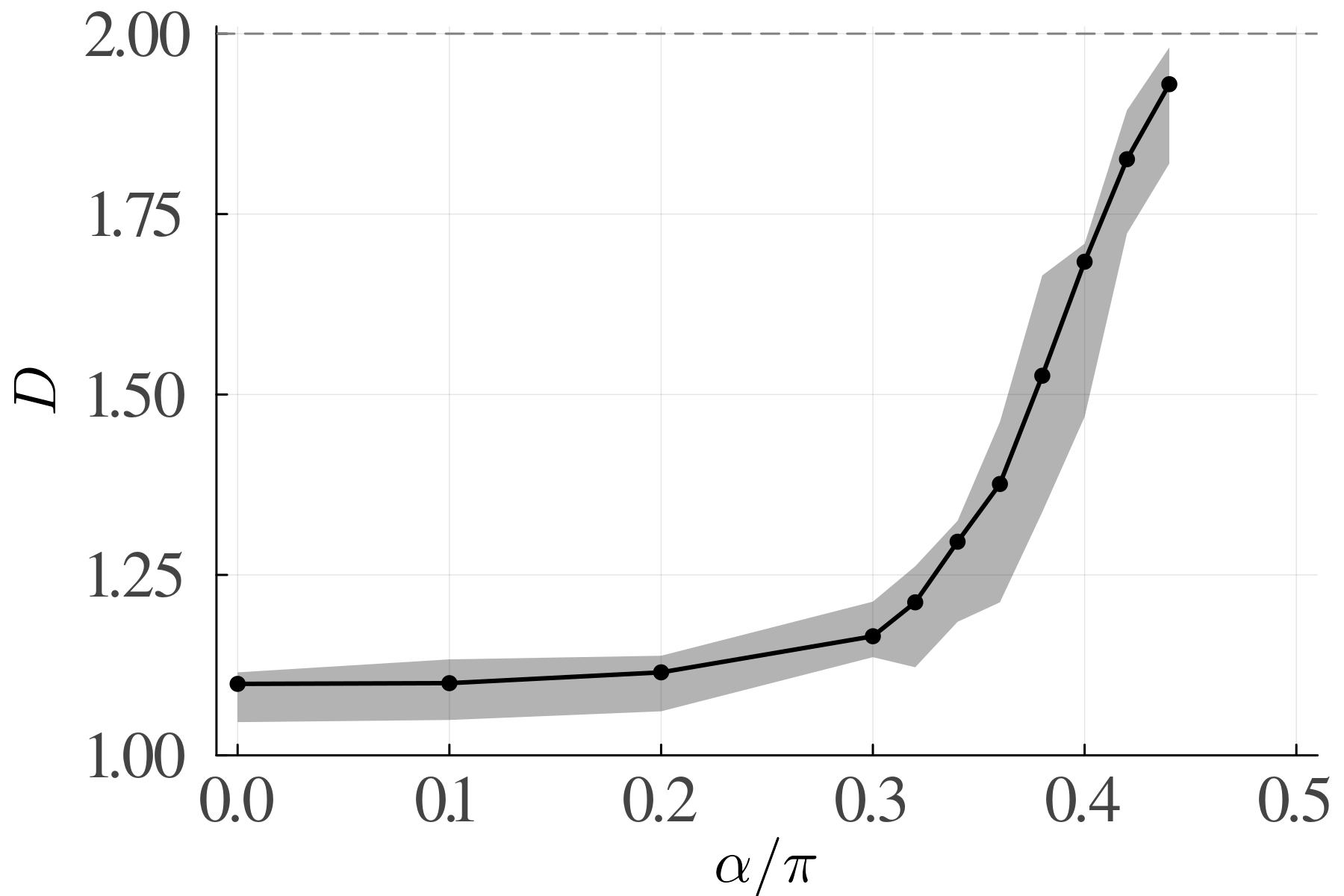}
\caption{Fractal dimension $D$ of the basin boundaries of the $q$-twisted states on the two-dimensional $(\epsilon_1, \epsilon_2)$-slice: the solid line shows the average over nine slices for $|q|<N/4$, while the gray shaded region represents the variation among different values of $q$; $N=80$.}
\label{fig:D}
\end{figure}

To characterize the fractal nature of the basins, we compute in Fig.~\ref{fig:D} the box-counting dimension $D$ \cite{strogatz2024nonlinear} of the basin boundaries of the $q$-twisted state, averaged over small $|q|$ (as their basin volumes are relatively large) and several two-dimensional random $(\epsilon_1, \epsilon_2)$-slices of the phase space. 
At $\alpha=0$, smooth basins in the full $N$-dimensional space appear fragmented in this two-dimensional slice, yielding a fractal dimension near $1$. The growth of $D$ is slow for $\alpha \in [0, 0.3\pi]$, followed by a sharp rise over $\alpha \in (0.3\pi, 0.4\pi]$. 
The system exhibits extreme sensitivity to initial conditions, not due to chaos, but rather a phenomenon known as final-state sensitivity \cite{grebogi1983final}. 
For $\alpha>0.4\pi$ we perform the numerical simulations with increased numerical accuracy \cite{note1} and obtain $D$ approaching $2$, i.e., the full dimension of the slice, as $\alpha \to \frac{\pi}{2}$. 
We assume that $D$ is insensitive to $q$ with $|q| < \frac{N}{4}$ (the gray region in Fig.~\ref{fig:D}). 
In summary, as $\alpha$ increases, the fractal dimension of the basin boundaries on two-dimensional slices grows from $1$ to $2$, with a sharp rise for $\alpha \in (0.3\pi, 0.44\pi]$. 
The approach of the basin boundary dimension to that of the slice suggests that the basin boundary appears riddled as $\alpha \to \frac{\pi}{2}$. Using an edge-tracking algorithm \cite{Attractors.jl}, we determine the edge states associated with the basin boundaries for small $\alpha \geq 0$, corresponding to the relatively smooth basin structure shown in Fig.~\ref{fig:basins-delpi}. These basin boundaries are formed by the stable manifolds of a family of unstable fixed points with a single unstable direction. At $\alpha=0$, these states are called {\it 1-saddles with winding number $q$}, given by \cite{delabays2017size}
\begin{equation}
\theta_j = \begin{cases} (j-1)\Delta_q', &\quad \text{if $1 \leq j < k$},\\ (j-3)\Delta_q' + \pi, &\quad \text{if $k \leq j \leq N$}, \end{cases}
\label{eq:1saddle}
\end{equation}
where $\Delta_q' := \frac{2q-1}{N-2}\pi$. The only unstable direction is associated with the jump at the $k$th oscillator. 
As $\alpha$ increases, this family of solutions deforms and undergoes a first bifurcation around $\alpha \in [0.24\pi, 0.31\pi]$, after which the states acquire additional unstable directions (see \hyperlink{SM:edge-tracking}{End Matter} for details). 
For larger $\alpha$, we conjecture that a chaotic saddle emerges, giving rise to fractal basin boundaries.

\begin{figure}[H]
\centering 
\includegraphics[width=\linewidth]{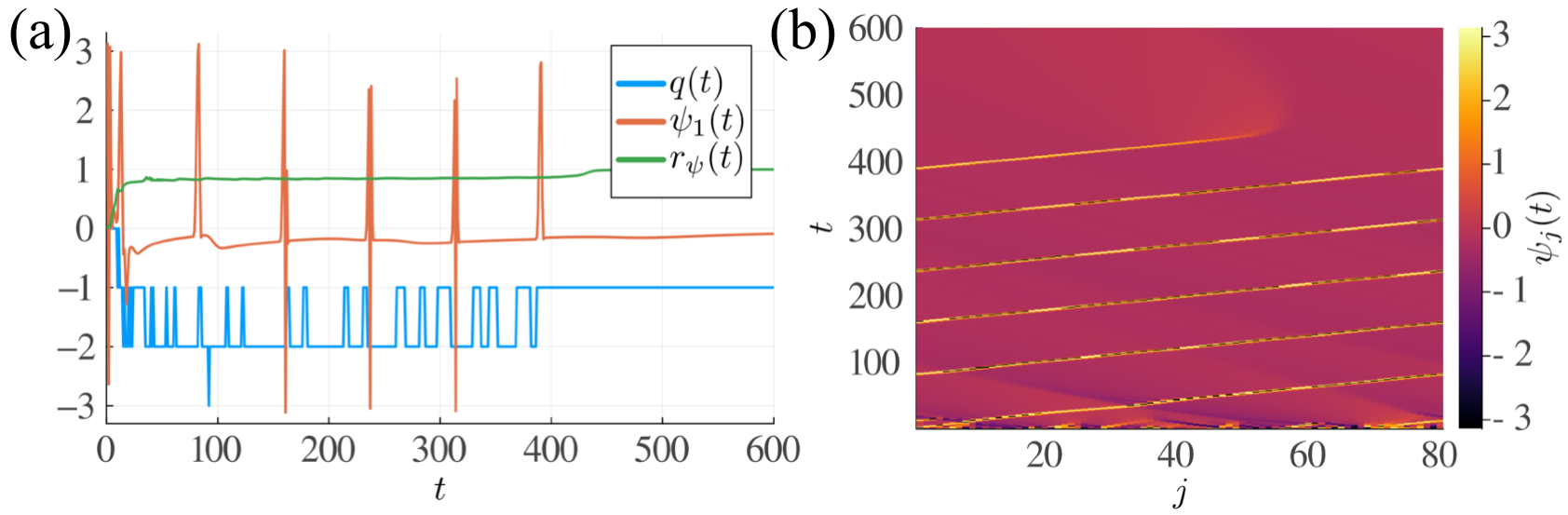}
\caption{(a) Typical time series of $\psi_1(t)$, corresponding winding number $q(t)$, order parameter $r_{\psi}(t)$ and (b) corresponding spatiotemporal evolution of $\boldsymbol{\psi}(t)$; here $\alpha=0.4\pi$, a random initial state and $N=80$.}
\label{fig:series}
\end{figure}

\textit{3.~Transient time.---} 
As the basin boundaries become increasingly fractal, the transient time required to reach a $q$-twisted state increases dramatically. We characterize this transient by the stabilization time $t_s$, defined as the first time at which the winding number $q(t):=\frac{1}{2\pi}\sum_{j=1}^{N}\psi_j(t)$ becomes constant \cite{groisman2025size}.
A typical time series of $q(t)$ is shown in Fig.~\ref{fig:series}(a) in blue, in comparison to the variable $\psi_1(t)$ and the Kuramoto order parameter $r_{\psi}(t) = \frac{1}{N}|\sum_{j=1}^N e^{i\psi_j(t)}|$. 

It has been shown that $t_s \propto \log(N)$ at $\alpha=0$ \cite{groisman2025size}, within which only short-range correlations are developed and a central limit theorem applies; consequently,  the basin size distribution of the $q$-twisted states follows a Gaussian at $\alpha=0$. 
We numerically found that even for $\alpha>0$, only short-range correlations can be developed, supporting the observation of the persistence of the Gaussian distributions in $q$ \cite{kim2026multistability}. For large $\alpha$, the transient dynamics often exhibit localized soliton-like waves on a twisted-state background, as shown in Fig.~\ref{fig:series}(b), which becomes an exact soliton in the continuum limit with $\alpha\to\frac{\pi}{2}$ \cite{rosenau2005phase}. This suggests the existence of an unstable set responsible for soliton-like waves. 
Trajectories can spend a long time shadowing this set before escaping to an attractor.

\begin{figure}[H]
\centering 
\includegraphics[width=\linewidth]{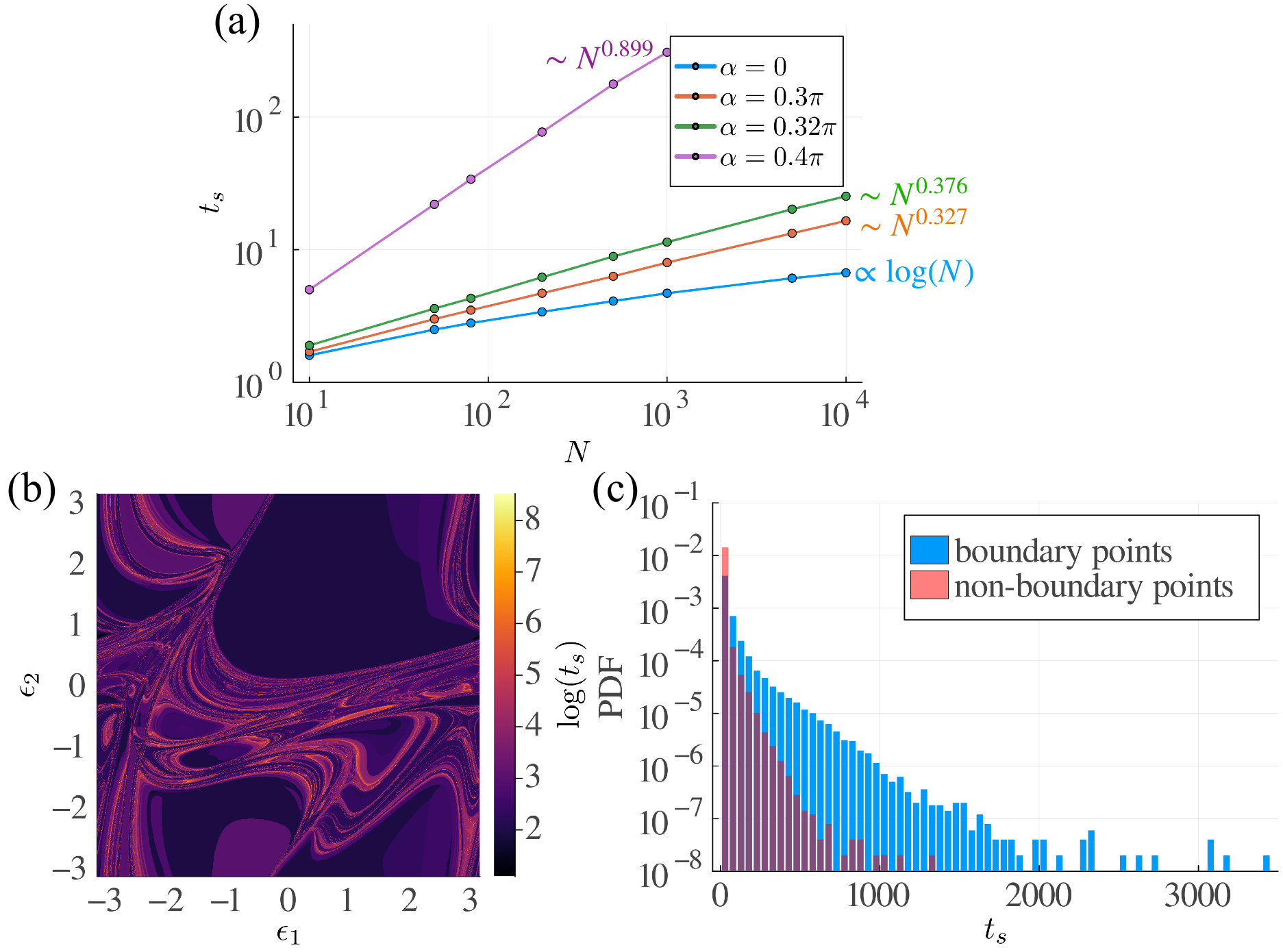}
\caption{(a) Averaged winding-number stabilization time $t_s$ as a function of system size $N$ for various $\alpha$ in log-log scale, where each point is averaged over $10^5$ random trajectories; (b) winding-number stabilization time $t_s$ for initial points on a $(\epsilon_1, \epsilon_2)$-slice at $\alpha=0.4\pi$ (cf.~Fig.~\ref{fig:basins-delpi}(d)); (c) distribution of $t_s$ separating basins boundary and non-boundary points, in semi-log scale.}
\label{fig:ts-ttran}
\end{figure}

Figure \ref{fig:ts-ttran}(a) shows that the scaling of the average $t_s$ with system size $N$ grows as $\alpha$ increases, ranging from logarithmic to a power-law. Up to $\alpha=0.4\pi$ the scaling remains sub-linear. 
Given the tendency of the quasi-riddled nature of the basins, we conjecture that the scaling will become super-linear, or even exponential -- a phenomenon known as supertransient  \cite{crutchfield1988attractors} as $\alpha \to \frac{\pi}{2}$. Fig.~\ref{fig:ts-ttran} (b) illustrates the stabilization time $t_s$  on the $(\epsilon_1, \epsilon_2)$-slice at $\alpha=0.4\pi$ (corresponding to the intermediate $\alpha$ represented in Fig.~\ref{fig:basins-delpi}(d)). In Fig.~\ref{fig:ts-ttran}(c), 
one finds that longer transient times $t_s$
occur near the basin boundary, whereas interior regions exhibit comparatively shorter $t_s$. 

\textit{4.~Outlook.---}
We have studied the basin metamorphosis of the twisted states in coupled phase oscillators. 
In the absence of a phase shift, each basin exhibits a simple octopus-shaped geometry, consisting of a head and tentacles \cite{zhang2021basins}. 
As a common phase shift $\alpha$ increases, our numerical results show that the basin boundaries develop fractal structure up to the limit $\alpha\to\frac{\pi}{2}$; approaching this limit, the basins appear riddled, and the dynamics becomes approximately volume-preserving. We have further investigated transient dynamics. The winding-number stabilization time scales as a power-law with system size, where the exponent grows with the phase shift. We have also examined the distribution of transient times and found that long transient times occur near the basin boundaries, whereas interior regions exhibit comparatively shorter transient times.

In summary: 
(1) at $\alpha=0$, the system exhibits gradient dynamics. (2) For small $\alpha$, the system remains close to gradient dynamics. The basins are relatively simple, with boundaries defined by the invariant manifolds of the index-1 saddle states, and exhibit short transient times. (3) For intermediate values of $\alpha$, fractal basin boundaries emerge, leading to longer transient times. 
(4) As $\alpha \to \frac{\pi}{2}$, the system approaches volume-preserving dynamics, and the basin boundaries become apparently riddled. In this regime, trapping by soliton-like waves becomes the dominant mechanism responsible for the prolonged transient times. (5) At $\alpha=\frac{\pi}{2}$, the system becomes Hamiltonian dynamics where solitons and kinks emerge in the continuum limit. These observations raise several open questions concerning the persistence of soliton-like waves under phase-shift variations, the geometry of their invariant manifolds, and the emergence of fractal basin boundaries, which we leave for future work.

\bibliography{ref}

@article{wiley2006size,
  title={The size of the sync basin},
  author={Wiley, Daniel A and Strogatz, Steven H and Girvan, Michelle},
  journal={Chaos},
  volume={16},
  pages={015103},
  year={2006},
  url={https://doi.org/10.1063/1.2165594}
}

@article{zhang2021basins,
  title={Basins with tentacles},
  author={Zhang, Yuanzhao and Strogatz, Steven H},
  journal={Phys. Rev. Lett.},
  volume={127},
  pages={194101},
  year={2021},
  url={https://doi.org/10.1103/PhysRevLett.127.194101}
}

@article{groisman2025size,
  title={Size of the sync basin resolved},
  author={Groisman, Pablo and De Vita, Cecilia and Bonder, Juli{\'a}n Fern{\'a}ndez and Zhang, Yuanzhao},
  journal={Phys. Rev. E},
  volume={112},
  pages={L052201},
  year={2025},
  url={https://doi.org/10.1103/pb4f-kqzq}
}

@article{sakaguchi1986soluble,
  title={A soluble active rotater model showing phase transitions via mutual entertainment},
  author={Sakaguchi, Hidetsugu and Kuramoto, Yoshiki},
  journal={Progr. of Theor. Phys.},
  volume={76},
  pages={576--581},
  year={1986},
  url={https://doi.org/10.1143/PTP.76.576}
}

@article{kuramoto2002coexistence,
  title={Coexistence of coherence and incoherence in nonlocally coupled phase oscillators},
  author={Kuramoto, Yoshiki and Battogtokh, Dorjsuren},
  journal={Nonlin. Phenom. Comp. Syst.},
  volume={5},
  pages={380--385},
  year={2002}, 
  url={http://www.j-npcs.org/abstracts/vol2002/v5no4/v5no4p380.html}
}

@article{alexander1992riddled,
  title={Riddled basins},
  author={Alexander, JC and Yorke, James A and You, Zhiping and Kan, Ittai},
  journal={Int. J. Bifurc. Chaos},
  volume={2},
  pages={795--813},
  year={1992},
  url={https://doi.org/10.1142/S0218127492000446}
}

@article{ott1993scaling,
  title={Scaling behavior of chaotic systems with riddled basins},
  author={Ott, Edward and Sommerer, John C and Alexander, James C and Kan, Ittai and Yorke, James A},
  journal={Phys. Rev. Lett.},
  volume={71},
  pages={4134},
  year={1993},
  url={https://doi.org/10.1103/PhysRevLett.71.4134}
}

@article{ashwin2000riddling,
  title={On riddling and weak attractors},
  author={Ashwin, Peter and Terry, John R},
  journal={Physica D},
  volume={142},
  pages={87--100},
  year={2000},
  url={https://doi.org/10.1016/S0167-2789(00)00062-2}
}

@article{aguirre2009fractal,
  title={Fractal structures in nonlinear dynamics},
  author={Aguirre, Jacobo and Viana, Ricardo L and Sanju{\'a}n, Miguel AF},
  journal={Rev. Mod. Phys.},
  volume={81},
  pages={333--386},
  year={2009},
  url={https://doi.org/10.1103/RevModPhys.81.333}
}

@article{dos2020basin,
  title={Basin of attraction for chimera states in a network of R{\"o}ssler oscillators},
  author={Dos Santos, Vagner and Borges, Fernando S and Iarosz, Kelly C and Caldas, Iber{\^e} L and Szezech, Jose Danilo and Viana, Ricardo L and Baptista, Murilo S and Batista, Antonio M},
  journal={Chaos},
  volume={30},
  pages={083115},
  year={2020},
  url={https://doi.org/10.1063/5.0014013}
}

@article{czajkowski2024riddled,
  title={Riddled basins of chaotic synchronization and unstable dimension variability in coupled Lorenz-like systems},
  author={Czajkowski, Bruno M and Viana, Ricardo L},
  journal={Chaos},
  volume={34},
  pages={093113},
  year={2024},
  url={https://doi.org/10.1063/5.0219961}
}

@article{rosenau2005phase,
  title={Phase compactons in chains of dispersively coupled oscillators},
  author={Rosenau, Philip and Pikovsky, Arkady},
  journal={Phys. Rev. Lett.},
  volume={94},
  pages={174102},
  year={2005},
  url={https://doi.org/10.1103/PhysRevLett.94.174102}
}

@article{ahnert2008traveling,
  title={Traveling waves and compactons in phase oscillator lattices},
  author={Ahnert, Karsten and Pikovsky, Arkady},
  journal={Chaos},
  volume={18},
  pages={037118},
  year={2008},
  url={https://doi.org/10.1063/1.2955758}
}

@article{smirnov2018solitary,
  title={Solitary synchronization waves in distributed oscillator populations},
  author={Smirnov, LA and Osipov, GV and Pikovsky, Arkady},
  journal={Phys. Rev. E},
  volume={98},
  pages={062222},
  year={2018},
  url={https://doi.org/10.1103/PhysRevE.98.062222}
}

@article{medeiros2018boundaries,
  title={Boundaries of Synchronization in Oscillators Networks},
  author={Medeiros, Everton S and Medrano-T, Rene O and Caldas, Iber{\^e} L and Feudel, Ulrike},
  journal={Phys. Rev. E},
  volume={98},
  pages={030201(R)},
  year={2018},
  url={https://doi.org/10.1103/PhysRevE.98.030201}
}

@article{kim2026multistability,
  title={Multistability and Control in Ring Networks of Phase Oscillators with Frequency Heterogeneity and Phase Lag},
  author={Kim, Soomin and Kori, Hiroshi},
  journal={Journal of the Physical Society of Japan},
  volume={95},
  number={6},
  pages={064005},
  year={2026},
  publisher={The Physical Society of Japan}, 
  url={https://journals.jps.jp/doi/full/10.7566/JPSJ.95.064005}
}

@book{strogatz2024nonlinear,
  title={Nonlinear dynamics and chaos: with applications to physics, biology, chemistry, and engineering},
  author={Strogatz, Steven H},
  year={2024},
  publisher={Chapman and Hall/CRC}, 
  url={https://doi.org/10.1201/9780429398490}
}

@book{dijkstra2013nonlinear,
  title={Nonlinear climate dynamics},
  author={Dijkstra, Henk A},
  year={2013},
  publisher={Cambridge University Press}, 
  url={https://doi.org/10.1017/CBO9781139034135}
}

@book{pikovsky1985universal,
  title={Synchronization: A universal concept in nonlinear sciences},
  author={Pikovsky, Arkady and Rosenblum, Michael and Kurths, J{\"u}rgen},
  year={1985},
  publisher={Cambridge University Press, Princeton}, 
  url={https://doi.org/10.1017/CBO9780511755743}
}

@book{may2001stability,
  title={Stability and complexity in model ecosystems},
  author={May, Robert M},
  volume={6},
  year={2001},
  publisher={Princeton university press}, 
  url={https://doi.org/10.1515/9780691206912}
}

@article{eaves2025nonlinear,
  title={Nonlinear stability measures of synchronised states in a power grid model},
  author={Eaves, Tom S},
  journal={J. Nonlinear Sci.},
  volume={35},
  pages={46},
  year={2025},
  url={https://doi.org/10.1007/s00332-025-10139-6}
}

@article{wunderling2020basin,
  title={Basin stability and limit cycles in a conceptual model for climate tipping cascades},
  author={Wunderling, Nico and Gelbrecht, Maximilian and Winkelmann, Ricarda and Kurths, J{\"u}rgen and Donges, Jonathan F},
  journal={New J. Phys.},
  volume={22},
  pages={123031},
  year={2020},
  url={https://doi.org/10.1088/1367-2630/abc98a}
}

@article{bollt2023fractal,
  title={Fractal basins as a mechanism for the nimble brain},
  author={Bollt, Erik and Fish, Jeremie and Kumar, Anil and Roque dos Santos, Edmilson and Laurienti, Paul J},
  journal={Sci. Rep.},
  volume={13},
  pages={20860},
  year={2023},
  url={https://doi.org/10.1038/s41598-023-45664-5}
}

@article{le2025asymmetric,
  title={Asymmetric coupling of nonchaotic Rulkov neurons: Fractal attractors, quasimultistability, and final state sensitivity},
  author={Le, Brandon B},
  journal={Phys. Rev. E},
  volume={111},
  pages={034201},
  year={2025},
  url={https://doi.org/10.1103/PhysRevE.111.034201}
}

@article{feudel2023rate,
  title={Rate-induced tipping in ecosystems and climate: the role of unstable states, basin boundaries and transient dynamics},
  author={Feudel, Ulrike},
  journal={NPG},
  volume={30},
  pages={481--502},
  year={2023},
  url={https://doi.org/10.5194/npg-30-481-2023}
}

@article{crutchfield1988attractors,
  title={Are attractors relevant to turbulence?},
  author={Crutchfield, James P and Kaneko, Kunihiko},
  journal={Phys. Rev. Lett.},
  volume={60},
  pages={2715},
  year={1988},
  url={https://doi.org/10.1103/PhysRevLett.60.2715}
}

@misc{note1, 
note = {We use Julia package \href{https://docs.sciml.ai/DiffEqDocs/stable/}{DifferentialEquations.jl} to numerically integrate the system of ordinary differential equations. Due to fractal basin boundaries, the following numerical solvers are used: for $\alpha \in [0, 0.36\pi)$ Euler with time-step dt=0.01, for $\alpha \in [0.36\pi, 0.44\pi)$ RK4 (4th-order Runge-Kutta) with tolerances (abstol, reltol)=($10^{-6}, 10^{-8}$), and for $\alpha \in [0.44\pi, 0.5\pi)$ AutoVern7(Rodas4P) with (abstol, reltol)=($10^{-12}, 10^{-12}$).}
}

@article{grebogi1983final,
  title={Final state sensitivity: an obstruction to predictability},
  author={Grebogi, Celso and McDonald, Steven W and Ott, Edward and Yorke, James A},
  journal={Phys. Lett. A},
  volume={99},
  number={9},
  pages={415--418},
  year={1983},
  url={https://doi.org/10.1016/0375-9601(83)90945-3}
}

@article{delabays2017size,
  title={The size of the sync basin revisited},
  author={Delabays, Robin and Tyloo, Melvyn and Jacquod, Philippe},
  journal={Chaos},
  volume={27},
  number={10},
  pages={103109},
  year={2017},
  url={https://doi.org/10.1063/1.4986156}
}

@software{Attractors.jl,
  author       = {George Datseris and others},
  title        = {JuliaDynamics/Attractors.jl},
  month        = may,
  year         = 2026,
  publisher    = {Zenodo},
  version      = {v1.36.0},
  doi          = {10.5281/zenodo.20342690},
  url          = {https://doi.org}
}

@misc{veltz:hal-02902346,
  TITLE = {{BifurcationKit.jl}},
  AUTHOR = {Veltz, Romain},
  URL = {https://hal.archives-ouvertes.fr/hal-02902346},
  INSTITUTION = {{Inria Sophia-Antipolis}},
  YEAR = {2020},
  MONTH = Jul,
  HAL_ID = {hal-02902346},
  HAL_VERSION = {v1}
}

\clearpage
\twocolumngrid
\appendix 
\section*{End Matter}
\hypertarget{SM:edge-tracking}{} % creates the invisible anchor point
\addcontentsline{toc}{section}{End Matter} 

As the system size $N$ increases, the number of 1-saddle states also increases. To illustrate how edge-tracking identifies the unstable invariant set underlying the basin boundaries, we first consider the minimal system size exhibiting multistability, $N=5$, for which the $q$-twisted states with $q = 0, \pm 1$ are the stable solutions of \eqref{eq:sys-theta}. 

\begin{figure}[H]
\centering 
\includegraphics[width=\linewidth]{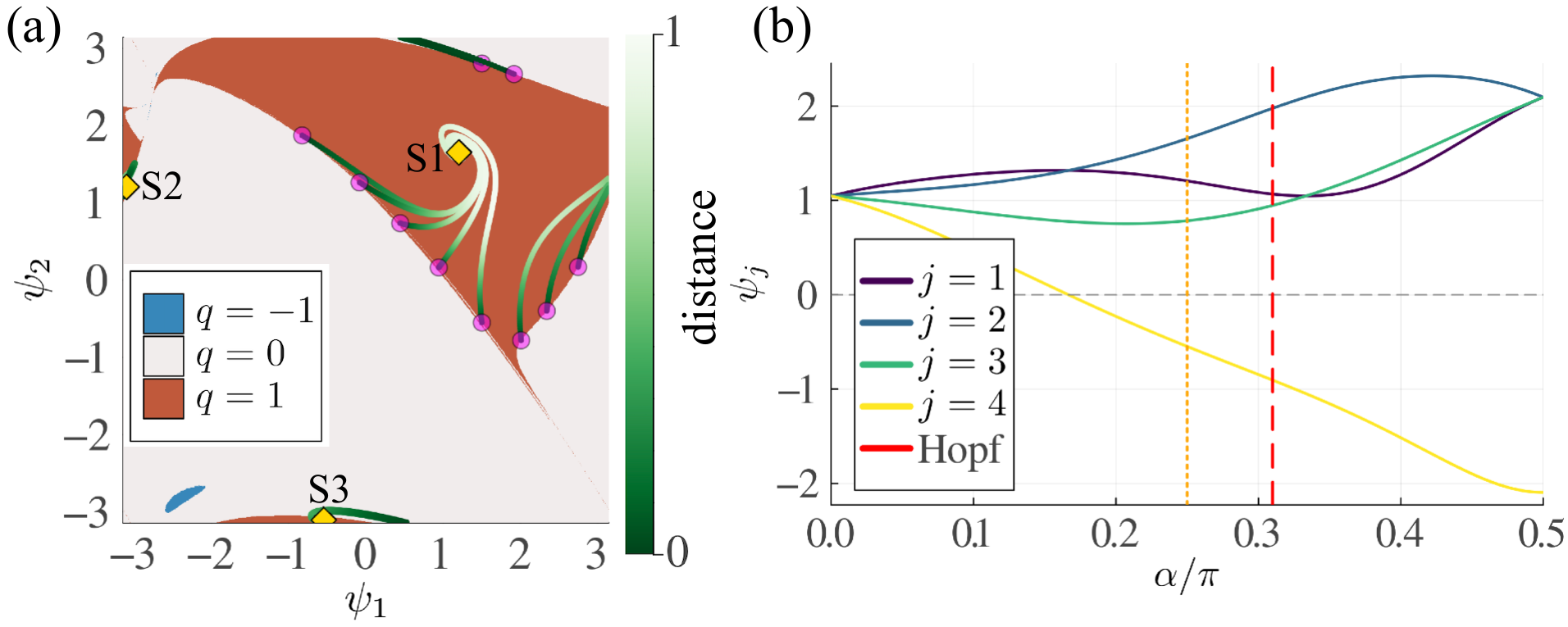}
\caption{$N=5$: (a) multiple edge-tracking locus converging to the unstable fixed points (S1, S2, S3) whose stable manifolds form the basin boundaries of the $q$-twisted states, here $\alpha=0.25\pi$; the green color gradient shows the normalized Euclidean distance from the displayed plane up to the converged point; these unstable fixed points are (b) continuations of the 1-saddle states with winding number $q=1$ (Eq.~\eqref{eq:1saddle}) in $\alpha \in [0, \frac{\pi}{2}]$, where the first bifurcation occurs at $\alpha \approx 0.31\pi$ (red dashed line); the orange dotted line highlights the coordinates of the point S1 in (a).}
\label{fig:edgestate}
\end{figure}

The edge-tracking algorithm implemented in the Julia package \texttt{Attractors.jl} \cite{Attractors.jl} starts from a pair of initial points belonging to different basins of attraction. In Fig.~\ref{fig:edgestate}(a), we visualize the edge-tracking process on the $(\psi_1, \psi_2)$-plane with $\psi_3=\psi_4=\psi_5=\Delta_1:=\frac{2\pi q}{N} = \frac{2\pi}{5}$, so that the $q=1$ twisted state, $\boldsymbol{\psi} = (\Delta_1, \Delta_1, \Delta_1, \Delta_1, \Delta_1)$, lies on this plane. We choose one initial point in the basin of the $q=0$ state (light region) and the other in the basin of the $q=1$ state (orange region). 

The algorithm first locates a point on the basin boundary (magenta dot), and then iteratively bisects to track along the boundary until it converges to an invariant saddle set, which in this case is a saddle fixed point (yellow diamond). 
The edge-tracking locus is shown in green-to-white color gradient indicating the normalized Euclidean distance from the displayed plane, where 0 means on the slice and 1 on the saddle. It is clear that the locus leaves this plane and moves deeper into the phase space. 

Repeating this process for multiple pairs of initial points, we find that all locus converge to one of the three fixed points labeled S1, S2 and S3. 
By extracting their coordinates and continuing them in $\alpha \in [0, \frac{\pi}{2}]$ using \texttt{BifurcationKit.jl} \cite{veltz:hal-02902346}, we identify them as continuations of the 1-saddles at $\alpha=0$. 
Specifically, for $N=5$ they are 1-saddles of $q=\pm 1$: (cf. Eq.~\eqref{eq:1saddle}, modulo translational symmetry)
$$
\boldsymbol{\psi}_{\text{1-saddle, $N$=$5$, $q$=$\pm 1$}} = (\Delta_{\pm 1}', \Delta_{\pm 1}', \Delta_{\pm 1}', \Delta_{\pm 1}', 2\pi - 4\Delta_{\pm 1}'), 
$$
where $\Delta_{\pm 1}' = \pm \frac{\pi}{3}$. 
Numerically, we find that the stable manifolds of the 1-saddles with $q=\pm 1$ form the basin boundaries of the $q=0$ and $q=\pm 1$ twisted states, while the basins of the $q=-1$ and $q=1$ twisted states do not appear to share a common boundary. Junction points where more than two basins meet may exist, but they seem to be rare for small $\alpha$. 

These saddle points undergo a Hopf bifurcation at $\alpha \approx 0.9736 \approx 0.31\pi$ (Fig.~\ref{fig:edgestate}(b)), reducing the dimension of their stable manifolds by two. 

The detailed continuation diagrams depend on the value of $N \bmod 4$, and are shown in Fig.~\ref{fig:1saddle-cont}: 
\begin{enumerate}[label={}, leftmargin=*]
\item Fig.~\ref{fig:1saddle-cont}(a): for $N \equiv 0 \pmod 4$ (e.g., $N=8, 80$), a fold bifurcation occurs, reducing the dimension of the stable manifold by one; 
\vspace{-0.5em}
\item Fig.~\ref{fig:1saddle-cont}(b): for $N \equiv 1 \pmod 4$ (e.g., $N=5, 81$), the 1-saddle branch first undergoes a Hopf bifurcation, followed by further bifurcations at larger $\alpha$, each increasing its instability; 
\vspace{-0.5em}
\item Fig.~\ref{fig:1saddle-cont}(c): for $N \equiv 2 \pmod 4$ (e.g., $N=6, 82$), similar to the case $N \equiv 0 \pmod 4$; 
\vspace{-0.5em}
\item Fig.~\ref{fig:1saddle-cont}(d): for $N \equiv 3 \pmod 4$ (e.g., $N=7, 83$), a similar fold bifurcation reduces the dimension of the stable manifold by one. 
\end{enumerate}

\begin{figure}[H]
\centering 
\includegraphics[width=\linewidth]{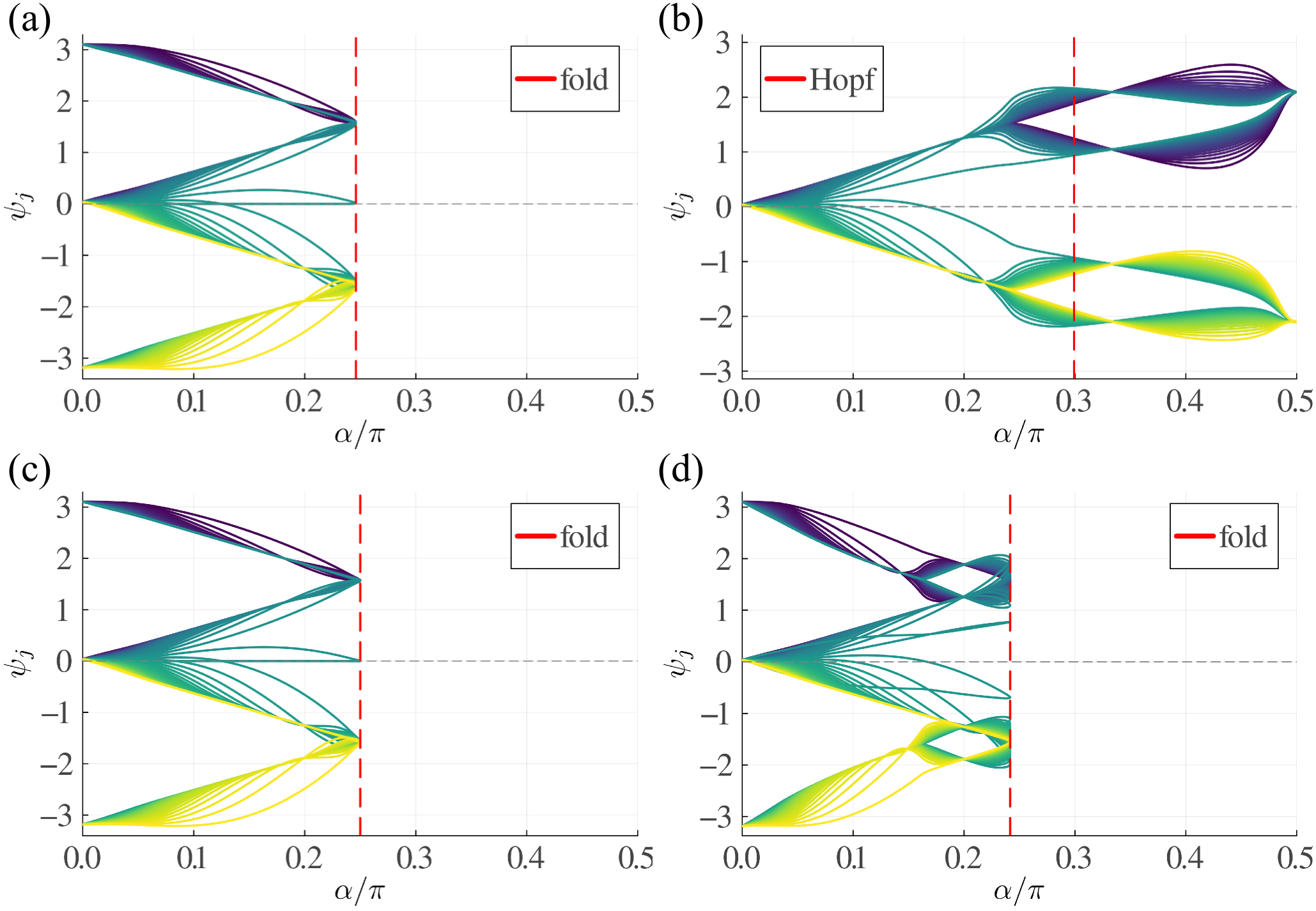}
\caption{Continuation of a 1-saddle state with winding number $q=1$ in $\alpha \in [0, \frac{\pi}{2}]$ for various system sizes: (a) $N=80$ or $N \equiv 0 \pmod 4$, (b) $N=81$ or $N \equiv 1 \pmod 4$, (c) $N=82$ or $N \equiv 2 \pmod 4$, and (d) $N=83$ or $N \equiv 3 \pmod 4$. The color gradient indicates the oscillator index, $j=1, ..., N-1$ from purple to yellow, and $\psi_N = 2\pi - \sum_{j=1}^{N-1}\psi_j$. As $\alpha$ increases from zero, only the first bifurcation is highlighted in red dashed line, which occurs at $\alpha \approx 0.25\pi, 0.3\pi, 0.25\pi, 0.24\pi$, from (a) to (d) respectively.}
\label{fig:1saddle-cont}
\end{figure}

Although the bifurcations differ among these four cases, a 1-saddle undergoes at least one bifurcation that increases its unstable direction, which occurs around $\alpha \in [0.24\pi, 0.31\pi]$. Consequently, the 1-saddles lose the codimension-1 stable manifolds required to form basin boundaries, and are not expected to remain the edge states for large values of $\alpha$.

\let\addcontentsline\oldaddcontentsline
	\cleardoublepage
	\onecolumngrid

\end{document}